\documentclass[%
 aip,
cp,  % Conference Proceedings
 amsmath,amssymb,%nobibnotes,
 reprint,%
]{revtex4-2}

\usepackage{graphicx}% Include figure files
\usepackage{dcolumn}% Align table columns on decimal point
\usepackage{bm}% bold math
\usepackage[utf8]{inputenc}
\usepackage[T1]{fontenc}
\usepackage{mathptmx} 

\def\mnras{MNRAS}
\def\apj{Astroph. J.}

\def\sovast{Sov. Astron.}

\begin{document}

\title{On the Origin of the Long Gamma-Ray Burst Afterglow as Synchrotron Radiation from Binary-Driven Hypernovae}% Force line breaks with \\

\author{Jorge A. Rueda} % Write as First name Surname
% \email[Corresponding author: ]{jorge.rueda@icra.it}
\email{jorge.rueda@icra.it}
%\author{Jorge A. Rueda}%
% \email{second.author@institution.edu.}
\affiliation{
ICRANet, Piazza della Repubblica 10, I-65122 Pescara, Italy\\
ICRA, Dipartimento di Fisica, Sapienza Universit\`a  di Roma, Piazzale Aldo Moro 5, I-00185 Roma, Italy\\
ICRANet-Ferrara, Dipartimento di Fisica e Scienze della Terra, Universit\`a degli Studi di Ferrara, Via Saragat 1, I--44122 Ferrara, Italy\\
Dipartimento di Fisica e Scienze della Terra, Universit\`a degli Studi di Ferrara, Via Saragat 1, I--44122 Ferrara, Italy\\
INAF, Istituto di Astrofisica e Planetologia Spaziali, Via Fosso del Cavaliere 100, 00133 Rome, Italy
% Force line breaks with \\ if necessary
}

% \author{Another's Name}
%  \email{third.author@anotherinstitution.edu}
% \affiliation{%
% Second institution and/or address% Force line breaks with \\ if necessary
% }%
% \affiliation{You would list an author's second affiliation (if applicable) here.}

\date{\today} % It is always \today, today, but any date may be explicitly specified
              % Not printed for conference proceedings

\begin{abstract}
Long gamma-ray bursts show an afterglow emission in the X-rays, optical, and radio wavelengths with luminosities that fade with time with a nearly identical power-law behavior. In this talk, I present an analytic treatment that shows that this afterglow is produced by synchrotron radiation from the supernova ejecta associated with binary-driven hypernovae.
\end{abstract}

\maketitle

%%%%%%%%%%%%%%%%%%%%%%%%%%%%%%%%%%%%%%%%%%%%%%%%%%%%%%
%%%%%%%%%%%%%%%%%%%%%%%%%%%%%%%%%%%%%%%%%%%%%%%%%%%%%%
\section{Introduction}\label{sec:1}
%%%%%%%%%%%%%%%%%%%%%%%%%%%%%%%%%%%%%%%%%%%%%%%%%%%%%%
%%%%%%%%%%%%%%%%%%%%%%%%%%%%%%%%%%%%%%%%%%%%%%%%%%%%%%

The \textit{Italian-Korean Symposium on Relativistic Astrophysics} has traditionally been an ideal place to exchange and present new ideas and progress on the latest results in our field of research. The fruitful atmosphere of academic exchange of the meeting has consistently stimulated novel theoretical developments. In this seventeenth version of the meeting (IK17), I present an analytic formulation of the synchrotron radiation originating in a binary-driven hypernova (BdHN), progenitors of long gamma-ray bursts (GRBs). Part of the contents of my talk at the IK 17 follow from my plenary talk at the \textit{16th Marcel Grossmann Meeting} (MG16), held the week before the IK17.

The synchrotron radiation in this model explains the distinct observational feature of the GRB afterglow: its luminosity fades with time with the same power-law behavior in the X-rays, optical, and radio wavelengths. The treatment here presented follows from a previous numerical model developed at ICRANet, already applied to the case of GRB 130427A \cite{2018ApJ...869..101R}, and also to GRBs 160509A, 160625B, 180728A, and 190114C \cite{2020ApJ...893..148R}. The aim has been to construct an analytic, self-consistent model that catches the physical situation at work and, in addition, allows a systematic analysis of the GRB afterglows. 

The presentation of the results in this article partially follows Rueda \textit{et al.} (to appear in IJMPD), the publication associated with my plenary talk at the MG16. First, I recall the BdHN scenario focusing on the properties that are essential for the development of the present theory of the afterglow. Then, I proceed to the concrete formulation that leads to the afterglow luminosity within the present model. Finally, I summarize the results and draw the conclusions.

%%%%%%%%%%%%%%%%%%%%%%%%%%%%%%%%%%%%%%%%%%%%%%%%%%%%%%
%%%%%%%%%%%%%%%%%%%%%%%%%%%%%%%%%%%%%%%%%%%%%%%%%%%%%%
\section{The binary-driven hypernova scenario}\label{sec:2}
%%%%%%%%%%%%%%%%%%%%%%%%%%%%%%%%%%%%%%%%%%%%%%%%%%%%%%
%%%%%%%%%%%%%%%%%%%%%%%%%%%%%%%%%%%%%%%%%%%%%%%%%%%%%%

BdHNe are progenitors of long GRBs (see \cite{2021MNRAS.504.5301R}, and references therein) characterized by a binary composed of a carbon-oxygen (CO) star and a neutron star (NS) companion in a tight orbit (orbital period of about 5 minutes). The core-collapse of the iron core of the CO star forms a new NS (hereafter $\nu$NS) at the SN center. The $\nu$NS formation process develops a strong shockwave that expands and injects energy into the CO star outer layers. This process unbinds the outer layers leading to a supernova (SN). The matter ejected by the SN partially accretes onto the NS companion and the $\nu$NS via matter fallback. The high temperature of the NS surface produces electron-positron pairs that annihilate into neutrino-antineutrinos. The high flux of neutrinos allows the accretion process to proceed at hypercritical (highly super-Eddington) rates \cite{2015ApJ...812..100B,2016ApJ...833..107B,2018ApJ...852..120B}. The fate of the NS companion strongly depends on the orbital period. I address the reader to \cite{2019ApJ...871...14B} for numerical simulations of the above process and a detailed analysis on the effect of the different binary parameters on the accretion onto the NS companion and onto the $\nu$NS.

In systems with short orbital periods of about 5 minutes, the hypercritical accretion brings the NS companion to the critical mass for gravitational collapse, forming a black hole (BH). This subclass of BdHN is called type I (BdHN I). A BdHN I leads to a new binary composed of a $\nu$NS and a BH. For orbital periods longer than about 5 minutes, the hypercritical accretion rate decreases, so the NS does not reach the critical mass. In this case, the accretion process leads to a more massive NS (MNS). This subclass is named BdHN of type II (BdHN II). A BdHN II leads to a new binary composed of a $\nu$NS and a massive NS. I refer the reader to Refs. \cite{2021IJMPD..3030007R,2021ARep...65.1026R,2019Univ....5..110R} for recent reviews on the BdHN scenario of long GRBs and the related physical phenomena.

There are two physical components of BdHN that are relevant in the afterglow emission. First, the expanding SN ejecta. Second, the $\nu$NS formed at the SN center. These components are common to BdHN I and II, so this theoretical model implies that the afterglows of these two systems must be very similar, as indeed is observed (see \cite{2020ApJ...893..148R} and references therein for details).

The $\nu$NS injects energy into the ejecta via energetic electrons. The magnetic field of the $\nu$NS remains relatively strong at large distances from the $\nu$NS where the ejecta expand, so the electrons in the ejecta lose their energy mainly by synchrotron radiation. This latter emission ultimately powers the afterglow of the GRB together with the pulsar emission of the $\nu$NS. Therefore, the explanation of the observed afterglow at the different wavelengths reveal crucial information on the $\nu$NS rotation rate, on the structure and strength of its magnetic field, and the SN ejecta expansion.

%%%%%%%%%%%%%%%%%%%%%%%%%%%%%%%%%%%%%%%%%%%%%%%%%%%%%%
%%%%%%%%%%%%%%%%%%%%%%%%%%%%%%%%%%%%%%%%%%%%%%%%%%%%%%
\section{Synchrotron radiation by the expanding supernova ejecta}\label{sec:3}
%%%%%%%%%%%%%%%%%%%%%%%%%%%%%%%%%%%%%%%%%%%%%%%%%%%%%%
%%%%%%%%%%%%%%%%%%%%%%%%%%%%%%%%%%%%%%%%%%%%%%%%%%%%%%

We start by modeling the expansion of the ejecta. For simplicity, we consider a spherically symmetric distribution extending at radii $r_i \in [R_*,R_{\rm max}]$, with corresponding velocities $v_i \in [v_*,v_{\rm max}]$, in self-similar expansion
\begin{equation}\label{eq:radius}
    r_i(t) = r_{i,0} \hat{t}^n,\qquad
    v_i(t) = n \frac{r_i(t)}{t} = v_{i,0} \hat{t}^{n-1},
\end{equation}
where $\hat{t} \equiv t/t_*$, being $t_* \equiv n R_{*,0}/v_{*,0}$. Here, $r_{i,0}$ and $v_{i,0}$ are the initial radius and velocity of the layer (so at times $t \ll t_*$ close to the beginning of the expansion. The case $n=1$ corresponds to a uniform expansion.

Electrons are continuously injected at some pace by the $\nu$NS into the ejecta. These electrons cool due to the expansion of the ejecta and by synchrotron radiation. The latter is due to the presence of the magnetic field of the $\nu$NS. We model the decrease of the magnetic field decreases with distance as a power-law of index $\mu$, so its strength at the layer position $r=r_i$ is
\begin{equation}\label{eq:B}
    B_i(t) = B_{i,0}\, \left[\frac{r_{i,0}}{r_i(t)}\right]^\mu = \frac{B_{i,0}}{\hat{t}^{\mu n}},
\end{equation}
where $B^{(0)}_i$ is the magnetic field strength at $r=r_{i,0}$. Because the electrons lose their energy very efficiently by synchrotron radiation (see details below), we can simplify our calculation by adopting that the radiation originates from the innermost layer of the ejecta, i.e. from $r=R_*$. Within this approximation, we do not need details of the density profile in the ejecta, and assume that the matter ahead is transparent to the radiation. The evolution of this layer, following Eqs. (\ref{eq:radius}), is given by $R_*(t) = R_{*,0} \hat{t}^n$, $v_*(t) = v_{*,0} \hat{t}^{n-1}$, $t_* = n R_{*,0}/v_{*,0}$, and from Eq. (\ref{eq:B}), this layer feels a magnetic field that decreases with time as $B_*(t) = B_{*,0} \hat{t}^{-\mu n}$.
 
The evolution of the distribution of radiating electrons per unit energy, $N(E,t)$, is determined by the kinetic equation accounting for the particle energy losses (see, e.g., Ref. \cite{1962SvA.....6..317K})
\begin{equation}\label{eq:kinetic}
    \frac{\partial N(E, t)}{\partial t}=-\frac{\partial}{\partial E}\left[\dot{E}\,N(E,t)\right] + Q(E,t),
\end{equation}
where $Q(E,t)$ is the number of injected electrons per unit time, per unit energy, and $\dot E$ is the electron energy loss rate. We follow the treatment presented in Ref. \cite{1973ApJ...186..249P} to solve the kinetic equation (\ref{eq:kinetic}). The solution of Eq. (\ref{eq:kinetic}) is given by
\begin{equation}\label{eq:Nsol}
    N(E,t) = \int_E^\infty Q[E_i, t_i(t,E_i,E)] \frac{\partial t_i}{\partial E} dE_i,
\end{equation}
where $E_i$ is the energy of the electron injected at time $t_i$. Therefore, to solve Eq. (\ref{eq:Nsol}) we have to set the injection rate $Q(E,t)$ and obtain the relation $t_i(t,E_i,E)$. The latter can be obtained from the solution of the energy balance equation accounting for adiabatic expansion and synchrotron radiation losses
\begin{equation}\label{eq:gammadot}
    -\dot E = \frac{E}{\tau_{\rm exp}} + P_{\rm syn}(E,t),
\end{equation}
where $P_{\rm syn}(E,t)$ is the bolometric synchrotron power
\begin{equation}\label{eq:Psynbol}
    P_{\rm syn}(E,t) = \beta B_*^2(t) E^2,
\end{equation}
with $\beta = 2e^4/(3 m_e^4 c^7)$, and
\begin{equation}\label{eq:tauexp}
    \tau_{\rm exp} \equiv \frac{R_*}{v_*} = \frac{t}{n} = \frac{t_*}{n}\hat{t},
\end{equation}
is the characteristic timescale of expansion. Equation~(\ref{eq:gammadot}) is a Riccati differential equation whose solution for the magnetic field (\ref{eq:B}) can be written as
\begin{equation}\label{eq:gammavst}
    E = \frac{E_i\,(t_i/t)^n}{1 + {\cal M} E_i t^n_i\left[ \frac{1}{\hat{t}_i^{n (1+2\mu)-1}} -  \frac{1}{\hat{t}^{n (1+2\mu)-1}}\right]},
\end{equation}
where we have introduced the constant
\begin{equation}\label{eq:M2}
    {\cal M}\equiv \frac{\beta B^2_{*,0} t_*^{1-n}}{n (1 + 2 \mu) - 1},
\end{equation}
which have units of $1/({\rm energy}\times {\rm time}^n)$. In the limit $t/t_* \gg 1$ and $n=1$, Eq. (\ref{eq:gammavst}) reduces to Eq. (3.3) of \cite{1973ApJ...186..249P}, and in the limit $t_* \to \infty$, reduces to the solution presented in Sec.~3 of \cite{1962SvA.....6..317K} for synchrotron losses in a constant magnetic field.

We turn to the distribution of the injected particles. We assume the following power-law function
\begin{equation}\label{eq:Q}
Q(E,t)=Q_0(t)E^{-\gamma}, \qquad 0\leq E \leq E_{\rm max},
\end{equation}
where $\gamma$ and $E_{\rm max}$ are parameters to be determined from the observational data, and $Q_0(t)$ can be related to the power released by the newborn central object and injected into the ejecta. We assume that the injected power has the form
\begin{equation}\label{eq:Lt}
L_{\rm inj}(t) = 
L_0 \left(1+\frac{t}{t_q}\right)^{-k},
\end{equation}
where $L_0$, $t_q$, and $k$ are model parameters. We have not chosen arbitrarily the functional form of Eq. (\ref{eq:Lt}), actually, both the power released by magnetic dipole braking and by fallback accretion obey this sort of time evolution.  Therefore, the function $Q_0(t)$ can be found from
\begin{equation}\label{eq:LandQ}
L_{\rm inj}(t) = \int_{0}^{E_{\rm max}} E\,Q(E,t) dE =  \int_{0}^{E_{\rm max}} Q_0(t) E^{1-\gamma} dE =Q_0(t)\frac{E_\mathrm{max}^{2-\gamma}}{2-\gamma},
\end{equation}
which using Eq.~(\ref{eq:Lt}) leads to
\begin{equation}\label{eq:Q0}
    Q_0(t) = q_0\left(1+\frac{t}{t_q}\right)^{-k},
\end{equation}
where $q_0 \equiv  (2-\gamma)L_0/E_{\rm max}^{2-\gamma}$.

We have specified the electron injection rate $Q(E,t)$ by Eqs. (\ref{eq:Q}) and (\ref{eq:Q0}), and the relation $t_i(t,E_i,E)$ is obtained by inverting Eq. (\ref{eq:gammavst}). We are thus ready to proceed to the integration of Eq. (\ref{eq:Nsol}). We can write $N(E,t)$ as a piecewise function of time and energy separating physical regimes in which simplifications and approximations allow to obtain an analytic solution of Eq. (\ref{eq:Nsol}). We anticipate that within the present model, the observational data of GRBs is contained at times $t<t_b$ and electron energies $E_b<E<E_{\rm max}$, where
\begin{equation}\label{eq:Eb}
    E_b = \frac{\hat{t}^{2 \mu n-1}}{{\cal M} t_*^n},\qquad t_b = t_* ({\cal M} t_*^n E_{\rm max})^{\frac{1}{2 \mu n-1}}.
\end{equation}
At these times and energies, synchrotron losses dominate and the solution of Eq. (\ref{eq:Nsol}) is well approximated by
\begin{align}\label{eq:N3}
&N(E,t)\approx \begin{cases}
    \frac{q_0}{\beta B_{*,0}^2 (\gamma-1)}\hat{t}^{2 \mu n} E^{-(\gamma+1)}, & t < t_q\\
   \frac{q_0}{\beta B_{*,0}^2 (\gamma-1)}\left(\frac{t_q}{t_*}\right)^{k}\hat{t}^{2 \mu n-k} E^{-(\gamma+1)}, &   t_q<t<t_b.
\end{cases}
\end{align}

With the knowledge of $N(E,t)$, we can proceed to estimate the synchrotron spectral density (energy per unit time, per unit frequency) from $J_{\rm syn}(\nu,t)d\nu =  j_{\rm syn}(\nu,E,t) N(E,t) dE$, where $j_{\rm syn}(\nu,E,t)$ is the synchrotron power radiated by an electron of energy $E$, at the radiation frequency $\nu$. Most of the synchrotron radiation is emitted in a narrow range of frequencies around the radiation \textit{critical frequency}, $\nu_{\rm crit}$. Thus, we can assume that each electron emits all the synchrotron radiation at
\begin{equation}\label{eq:nuc}
    \nu \approx \nu_{\rm crit} \approx \alpha B_* E^2,
\end{equation}
where $\alpha = 3 e/(4\pi m_e^3 c^5)$. This implies that we can directly relate the electron's energy to the radiation frequency, and $j_{\rm syn}(\nu,E,t)$ can be approximated to the bolometric power
\begin{equation}\label{eq:Psynbol2}
    j_{\rm syn}(\nu, E, t) \approx P_{\rm syn}(\nu,t) = \frac{\beta}{\alpha} B_* \nu,
\end{equation}
where we have replaced Eq. (\ref{eq:nuc}) into Eq. (\ref{eq:Psynbol}). Within this approximation, the synchrotron spectral density is
\begin{equation}\label{eq:Jnu1}
    J_{\rm syn}(\nu,t) \approx P_{\rm syn}(\nu,t) N(E,t) \frac{dE}{d\nu} = \frac{\beta}{2} \alpha^{\frac{p-3}{2}} \eta B_{*,0}^{\frac{p+1}{2}}\hat{t}^{\frac{2 l-\mu n(p+1)}{2}}\nu^{\frac{1-p}{2}},
\end{equation}
where we have used
\begin{equation}\label{eq:Ngeneric}
    N(E,t) = \eta\,\hat{t}^l E^{-p},
\end{equation}
being $\eta$, $l$, and $p$ known constants from Eq. (\ref{eq:N3}). 

The synchrotron luminosity radiated in the frequencies $[\nu_1,\nu_2]$ can be then obtained as
\begin{equation}\label{eq:Lnu}
    L_{\rm syn}(\nu_1,\nu_2; t) = \int_{\nu_1}^{\nu_2} J_{\rm syn}(\nu,t)d\nu \approx \nu J_{\rm syn}(\nu,t)= \frac{\beta}{2} \alpha^{\frac{p-3}{2}} \eta B_{*,0}^{\frac{p+1}{2}}\hat{t}^{\frac{2 l-\mu n(p+1)}{2}}\nu^{\frac{3-p}{2}},
\end{equation}
where $\nu_1=\nu$, $\nu_2=\nu+\Delta\nu$, we have made the approximation $\Delta\nu/\nu\ll 1$, and used Eq. (\ref{eq:Jnu1}).

%%%%%%%%%%%%%%%%%%%%%%%%%%%%%%%%%%%%%%%%%%%%%%%%%%%
%%%%%%%%%%%%%%%%%%%%%%%%%%%%%%%%%%%%%%%%%%%%%%%%%%%
\section{Evolution of the $\nu$NS and its pulsar emission}\label{sec:4}
%%%%%%%%%%%%%%%%%%%%%%%%%%%%%%%%%%%%%%%%%%%%%%%%%%%
%%%%%%%%%%%%%%%%%%%%%%%%%%%%%%%%%%%%%%%%%%%%%%%%%%%

The $\nu$NS, owing to its magnetic field, emits also pulsar-like radiation. We adopt a dipole+quadrupole magnetic field model \cite[see][for details]{2015MNRAS.450..714P}. The total spindown luminosity in this model is given by
\begin{equation}\label{eq:Lsd}
    L_{\rm sd} = L_{\rm dip} + L_{\rm quad} = \frac{2}{3 c^3} \Omega^4 B_{\rm dip}^2 R^6 \sin^2\chi_1 \left( 1 + \xi^2 \frac{16}{45} \frac{R^2 \Omega^2}{c^2} \right),
\end{equation}
where $\Omega$ and $R$ are the stellar angular velocity and radius, and $\xi$ defines the quadrupole to dipole magnetic field strength ratio
\begin{equation}\label{eq:eta}
    \xi \equiv \sqrt{\cos^2\chi_2+10\sin^2\chi_2} \frac{B_{\rm quad}}{B_{\rm dip}},
\end{equation}
in which the modes are separable in a straightforward fashion. For example, the $m = 0$ mode is represented by $\chi_1 = 0$ and any value of $\chi_2$, the $m = 1$ mode is given by $(\chi_1, \chi_2) = (90^\circ, 0^\circ)$, and the $m = 2$ mode by $(\chi_1, \chi_2) = (90^\circ, 90^\circ)$. 

The evolution of the $\nu$NS structure is driven by the energy balance equation
\begin{equation}\label{eq:Erot}
	-\dot{E} = -(\dot{W}+\dot{T}) = L_{\rm tot} = L_{\rm inj} + L_{\rm sd},
\end{equation}
where $W$ and $T$ are, respectively, the $\nu$NS gravitational and rotational energy. The solution of the differential equation (\ref{eq:Erot}) for the spindown luminosity (\ref{eq:Lsd}) gives the evolution with time of the rotation angular frequency $\Omega(t)$, so the time evolution of the $\nu$NS pulsar luminosity, $L_{\rm sd}(t)$.

%%%%%%%%%%%%%%%%%%%%%%%%%%%%%%%%%%%%%%%%%%%%%%%%%%%
%%%%%%%%%%%%%%%%%%%%%%%%%%%%%%%%%%%%%%%%%%%%%%%%%%%
\section{Conclusions}
%%%%%%%%%%%%%%%%%%%%%%%%%%%%%%%%%%%%%%%%%%%%%%%%%%%
%%%%%%%%%%%%%%%%%%%%%%%%%%%%%%%%%%%%%%%%%%%%%%%%%%%

The synchrotron radiation by electrons injected by the $\nu$NS into the SN ejecta in a BdHN produces a luminosity that follow a power-law behavior both in time and radiation frequency, see Eq. (\ref{eq:Lnu}), $L_{\rm syn} \propto t^{\frac{2 l -\mu n (p+1)}{2}} \nu^{\frac{3-p}{2}}$, where $p = \gamma+1$ and the value of $l$ depend on whether we are in the phase of constant injection or power-law injection (see Eq. \ref{eq:N3} for details). The constant $n$ determines the law of expansion, Eq. (\ref{eq:radius}), and $\mu$ the radial dependence of the magnetic field strength, Eq. (\ref{eq:B}). Therefore, the synchrotron luminosity is characterized by the same power-law index in the X-rays, optical, and radio wavelengths, providing the system is in the same physical regime, namely the evolution is constrained at times $t<t_b$, and the electron energies are in the range $E_b<E<E_{\rm max}$, see Eq. (\ref{eq:Eb}). Otherwise, the synchrotron luminosity is characterized by different power-law behaviors determined by the dominance of adiabatic expansion losses over synchrotron losses. In our cases of study, we do not see evidence in the GRB afterglow data of the occurrence of such physical regime. This implies that the ratio of the luminosity at different frequencies depends only on the power-law index of the injection rate, namely
\begin{equation}\label{eq:Lratio}
    \frac{L_{\rm syn} (\nu_1)}{L_{\rm syn} (\nu_2)} = \left( \frac{\nu_1}{\nu_2} \right)^{\frac{3-p}{2}} = \left( \frac{\nu_1}{\nu_2} \right)^{\frac{2-\gamma}{2}}.
\end{equation}
The value of $\gamma$ can be set with the knowledge of the X to optical luminosity ratio. This, in turn, fixes automatically the optical to radio luminosity ratio. The fact that the single value of $\gamma$ gives the correct luminosity at different frequencies gives strong support to the model since it implies the correct description of the afterglow spectrum.

In due time, when the synchrotron luminosity has faded low enough, the pulsar emission of the $\nu$NS given by Eq. (\ref{eq:Lsd}) becomes observable in the X-rays. This occurs when $L_{\rm syn}(\nu_X,t) \sim L_{\rm sd}(t)$. The time of the pulsar emergence depends on the model parameters, specifically it gives information on the $\nu$NS rotation angular velocity at birth, and on the strength of the dipole and quadrupole components of the magnetic field.

The main physical picture of the present model is: electrons are injected by a newborn compact star into the matter expelled in the cataclysmic event and that expand at high velocity. These electrons penetrating the ejecta emit synchrotron emission because of the magnetic field of the compact star. These general features are also present in another kind of cataclysmic events, \textit{compact star binary mergers}. Therefore, the present model predicts that afterglows of similar properties might be also produced in binary neutron star mergers or binary white dwarf mergers (Rueda \textit{et al.}, to appear in IJMPD).

%%%%%%%%%%%%%%%%%%%%%%%%%%%%%%%%%%%%%%%%%%%%%%%%%%%%
%%%%%%%%%%%%%%%%%%%%%%%%%%%%%%%%%%%%%%%%%%%%%%%%%%%%
\begin{acknowledgments}
%%%%%%%%%%%%%%%%%%%%%%%%%%%%%%%%%%%%%%%%%%%%%%%%%%%%
%%%%%%%%%%%%%%%%%%%%%%%%%%%%%%%%%%%%%%%%%%%%%%%%%%%%

I thank the members of ICRANet Faculty Liang Li, Rahim Moradi, Remo Ruffini, Narek Sahakyan, Gregory Vereshchagin, and Yu Wang for discussions and help in testing the present analytic model in several observed GRB afterglows (presented elsewhere). I thank Sang Pyo Kim that shared with me, after my talk, his thoughts on the analytic solution of the Riccati equation given by the energy balance of electrons, and that confirmed the solution presented in the talk. Finally, I express my gratitude to the organizers of the conference and all the attendees for their engagement in keeping a constructive academic atmosphere.
\end{acknowledgments}

\providecommand{\noopsort}[1]{}\providecommand{\singleletter}[1]{#1}%

\end{document}